\documentstyle [11pt]{article}
\def\ZzZ{{\hbox{\tenrm Z\kern-.31em{Z}}}}
\def\CcC{{\hbox{\tenrm C\kern-.45em{\vrule height.67em width0.08em depth-
.04em \hskip.45em }}}}

\newcommand{\bc}{\begin{center}}
\newcommand{\ec}{\end{center}}
\newcommand{\be}{\begin{equation}}
\newcommand{\ee}{\end{equation}}
\newcommand{\bea}{\begin{eqnarray}}
\newcommand{\eea}{\end{eqnarray}}
\newcommand{\bs}{\begin{subequations}}
\newcommand{\es}{\end{subequations}}
\newcommand{\beq}{\begin{eqalignno}}
\newcommand{\eeq}{\end{eqalignno}}

\begin{document}

\bc {

{\bf The dissipative brain}

\bigskip
\bigskip

Giuseppe Vitiello

\medskip

Dipartimento di Fisica ``E.R.Caianiello", Universit\`a di
Salerno,\\
and INFM, Sezione di Salerno, 84100 Salerno, Italy

e-mail: vitiello@sa.infn.it }
\bigskip

\ec

I review the dissipative quantum model of brain and discuss its
recent developments related with the r\^ole of entanglement,
quantum noise and chaos. Some comments on consciousness in the
frame of the dissipative model are also presented. Dissipation
seems to account for the medial character of consciousness, for
its being in the present (the Now), its un-dividable unity, its
intrinsic subjectivity (autonomy). Finally, essential features of
a conscious artificial device, if ever one can construct it, are
briefly commented upon, also in relation to a device able to
exhibit mistakes in its behavior. The name I give to such a
hypothetical device is Spartacus.

\smallskip


\bigskip
\bigskip

\section{Introduction}
%

I have been always attracted by the unitary conception of
knowledge, which is ever present in some streams of our cultural
inheritance. Perhaps, one of the most vivid expressions of such a
comprehensive view of the world has been provided by Titus
Lucretius Carus in his {\it De rerum natura}. There are no
territories forbidden to our search and there are no separate
domains of knowledge,

\indent{''...we must not only give a correct account of celestial

matter, explaining in what way the wandering of the sun

and moon occur and by what power things happen on earth.

We must also take special care and employ keen reasoning

to see where the soul and the nature of the mind come from,...''}

(Titus Lucretius Carus, 99--55 B.C).

Therefore, searching in territories not traditionally explored by
physicists should not require special justifications.

In this paper I would like to review briefly the extension of the
quantum model of the brain to the dissipative dynamics. For sake
of brevity, I will mostly present results rather than their
derivations. The interested reader may find formal details in the
quoted literature. I will comment on consciousness in the
dissipative model frame and finally I will briefly discuss the
essential features which an artificial conscious device should
have, if ever it will be possible to construct it, and its
relation with a device able to exhibit mistakes in its behavior.

The extension of the quantum model to the dissipative dynamics is
required in order to solve the {\it overprinting} problem, namely
the fact that in the Ricciardi and Umezawa model (Ricciardi and
Umezawa, 1967) the memory capacity is extremely small: any
successive memory printing overwrites on the previously recorded
memory.

The proposed solution (Vitiello, 1995)  relies on two facts. One
is that the brain is a system permanently coupled with the
environment (an ``open'' or ``dissipative'' system). The other one
is a crucial property of quantum field theory (QFT), i.e. the
existence of infinitely many states of minimal energy, the so
called vacuum states or ground states. On each of these vacua
there can be built a full set (a space) of other states of nonzero
energy. We have thus infinitely many state spaces, which, in
technical words, are called ``representations of the canonical
commutation relations''. I will refer to the collection of all
these spaces or representations as the ``memory space''. The
vacuum of each of these spaces is characterized by a specific
ordering and is identified by its {\it code}, which is the value
of the ``order parameter'', the macroscopic observable
characterizing indeed the ordering present in that vacuum. Vacua
(or the corresponding spaces) identified by different codes are
``distinct'' vacua in the sense that one of them cannot be reduced
(transformed) into another one of them. In technical words, they
are ``unitary inequivalent vacua'' (or unitary inequivalent spaces
or representations).

In the dissipative quantum model of brain the vacuum code is taken
to be the memory code. A given memory is represented by a given
degree of ordering. A huge number of memory records can be thus
stored, each one in a vacuum of given code. In the original model
by Ricciardi and Umezawa only one vacuum is available for memory
printing. In the {\it dissipative} model all the vacua are
available for memory printing.

\bigskip
\bigskip
\section{Broken symmetry and order}

In the quantum model a crucial r\^{o}le is played by the mechanism
of ''the spontaneous breakdown of symmetry'' by which the
invariance (the symmetry) of the field equations {\it manifests}
itself into ordered patterns in the vacuum state. The symmetry is
said to be broken since the vacuum state does not possess the full
symmetry of the field equations (the dynamics). The ``order'' {\it
is} indeed such a ``lack of symmetry''.

One can show that when symmetry is broken the invariance of the
field equations implies the existence of quanta, the so called
Nambu-Goldstone (NG) quanta, which, propagating through the whole
system volume, are the carrier of the ordering information, they
are the long range {\it correlation modes}: in the crystal, for
example, the ordering information is the one specifying the
lattice arrangement.

The presence ({\it the condensation}) in the vacuum state of NG
quanta thus describes the ordering. When ordering is achieved,
each of the elementary components of the system is ``trapped'' in
a specific space-time behavior (e.g. a specific space position, a
specific oscillation mode, etc.). Ordering implies thus the
freezing of some of the degrees of freedom of the elementary
components. Or, in other words, their {\it coherent} motion or
behavior. It is such a coherent, collective behavior that
macroscopically manifests itself as the ordered pattern: the
microscopic quantum behavior thus provides macroscopic
(collective) properties. We have a ``change of scale'', from
microscopic to macroscopic, and the ordered state is called a {\it
macroscopic quantum state}. NG quanta are  therefore also called
collective modes. In the dissipative  model of brain these NG
quanta are called ``dipole wave quanta'' (dwq) since they
originate from the breakdown of the electrical dipole rotational
symmetry (Vitiello, 1995; Del Giudice et al., 1985, 1986; Jibu and
Yasue, 1995).

In order for the NG quanta to be able to span the full system
volume and thus set up the ordered pattern, their mass has to be
zero (or quasi--zero in realistic conditions of finite system
sizes). In their lowest momentum state NG quanta thus do not carry
energy. For this reason, the vacuum state where even a very large
number of them is condensed is a state of minimal energy. This
guaranties the stability of the ordered vacuum, namely of the
memory record in the quantum model of brain.

Incidentally, I observe that in this model the variables are basic
quantum field variables (the electrical dipole field). In the
quantum model ``we do not intend", Ricciardi and Umezawa say ``to
consider necessarily the neurons as the fundamental units of the
brain". Moreover,  Stuart, Takahashi and Umezawa (1978)
have also remarked that ``it is difficult to consider neurons as
quantum objects''.

In principle, any of the ordered patterns compatible with the
invariance of the field equations can be realized in the process
of symmetry breaking. This is why symmetry breaking is said to be
``spontaneous''. The point is that the ordered pattern which is
actually realized is the output of the system {\it inner}
dynamics. The process of symmetry breaking is triggered by some
external input; the ``choice'' of the specific symmetry pattern
which is actually realized is, on the contrary, ``internal'' to
system. Therefore one speaks of {\it self-organizing} dynamics:
ordering is an inner (spontaneous, indeed) dynamical process. This
feature of spontaneous symmetry breaking is common to solid state
physics and high energy physics, and it is of particular interest
when modelling the brain: in the brain, contrary to the computer
case, ordering is not imported from the outside, it is the
outgrowth of an ``internal'' dynamical process of the system.

The generation of ordered, coherent patterns is thus the dynamical
result of the system elementary component interactions.

\bigskip
\bigskip
\section{The brain is a dissipative system}

In the quantum model of brain a specific memory is associated to a
specific degree of ordering (a specific value of the vacuum {\it
code}). The overprinting problem then reduces to the problem of
making available all possible vacua, or, in other words, to attach
a specific label (the code value) to a given vacuum under the
trigger of a specific external input.

On the other hand, it is evident that in its continual interaction
with the environment the brain's time evolution is irreversible,
i.e., technically speaking, it is {\it non}--unitary. Getting
information from the outside world, which is a feature
characterizing the brain, makes the brain dynamics intrinsically
irreversible. I have elsewhere depicted such a situation by
mentioning the way of saying ...{\it Now you know it!}..., which
indeed means that once one gets some information, he/she is
nevermore the same person as before. Getting information
introduces the {\it Now} in our experience, or, in different
words, the feeling of the past and of the future, of the arrow of
time pointing forward in time. Without getting information there
would be neither forward nor backward in time. However, we {\it
cannot avoid} getting information, being opened on the world
(including our inner world, ourself). The brain is an open,
dissipative system. The brain closed on the world is a dead brain,
physiology tells us. Isolation of the brain (closure to the world)
produces serious pathologies. Thus, the extension of the quantum
model of brain to the dissipative dynamics appears to be a
necessity (Vitiello, 1995).

Then, the mathematical formalism for quantum dissipation
(Celeghini, Rasetti and Vitiello, 1992) {\it requires} the
doubling of the brain degrees of freedom. The doubled degrees of
freedom, say $\tilde A$ (the tilde quanta; the non--tilde quanta
$A$ denoting the brain degrees of freedom), are meant to represent
the environment to which the brain is coupled. The physical
meaning of the doubling is the one of ensuring the balance of the
energy flux between the system and the environment.

The environment thus represented by the doubled degrees of freedom
appears described as the ``time--reversed copy'' (the {\it
Double}) of the brain. The environment is ``modelled'' on the
brain. Time--reversed since the energy flux outgoing from the
brain is incoming into the environment, and vice versa.

In addition, the quantum dissipation formalism implies that the
full operator describing the system time evolution includes the
operator describing the coupling between the non--tilde and the
tilde quanta. At the same time, such a coupling term acts as the
mathematical tool to attach the label to the vacua (and thus to
distinguish among different memories). This label is
time--dependent: the system states are thus time--dependent
states.

In this new light, the time evolution operator is readily
recognized to be the ``free energy'' operator, not just the
Hamiltonian operator, as it would be in the absence of
dissipation. The free energy operator is made indeed by the
Hamiltonian operator, which controls the reversible (unitary) part
of time evolution, plus the non--tilde/tilde coupling term, which
is recognized to be proportional to the {\it entropy} operator and
controls indeed the non-unitary, irreversible part of time
evolution. In a thermodynamical language this last term describes
the heat term in the system energy.

The doubling of the degrees of freedom in the dissipative model
thus arises as a consequence of the irreversible time evolution.

Once thermal aspects in the dissipative model have been also
recognized, the memory state is found (Vitiello, 1995) to be a
{\it non--equilibrium} Thermo Field Dynamics (TFD) state. TFD is
the QFT formalism for thermal systems introduced by Takahashi and
Umezawa (Takahashi and Umezawa, 1975; Umezawa, 1993) which
provides an explicit representation of the so called
Gelfand--Naimark--Segal (GNS) construction in the $C^{*}$--algebra
formalism (Ojima, 1981). TFD was not devised for the study of the
brain, but for the study of solid state physics, to which it has
been successfully applied.

In equilibrium TFD the system time evolution is fully controlled
by the Hamiltonian. The operator necessary to attach the label to
the thermal states (the label is temperature in that case) is not
a term of the time evolution operator (as, on the contrary, in the
dissipative model). Non-equilibrium transitions (non-unitary time
evolution) in thermal systems have been considered later on in the
time-dependent TFD formalism (Umezawa, 1993; and references
therein quoted). The non--equilibrium character of the brain
dynamics makes the dissipative model substantially different from
equilibrium TFD.

One can show that the dynamics now includes, when the system
volume is large but finite, the possibility of transitions {\it
through} inequivalent (labelled by different codes) vacua: in this
way, at once, familiar phenomena such as memory associations,
memory confusion, even the possibility to forget some memories, or
else difficulties in recovering memory, are described by the
dissipative model. The dissipative character of the dynamics thus
accounts for many features of the brain behavior and for its huge
memory capacity: now, indeed, all the differently coded vacua
become accessible to the memory printing process. Their unitary
inequivalence at large volume guaranties protection from
overprinting, not excluding, however, due to realistic boundary
effects, the processes of association, confusion, etc. just
mentioned.

The general scheme of the dissipative quantum model can be
summarized as follows. The starting point is that the brain is
permanently coupled to the environment. Of course, the specific
details of such a coupling may be very intricate and changeable so
that they are difficult to be measured and known. One possible
strategy is to average the effects of the coupling and represent
them, at some degree of accuracy, by means of some ``effective''
interaction. Another possibility is to take into account the
environmental influence on the brain by a suitable {\it choice}
of the brain vacuum%
\index{vacuum state} state.  Such a choice is triggered by the
external input (breakdown of the symmetry), and it actually is the
end point of the internal (spontaneous) dynamical process of the
brain (self-organization). The chosen vacuum thus carries the {\it
signature} (memory) of the reciprocal brain--environment influence
at a given time under given boundary conditions. A change in the
brain--environment reciprocal influence then would correspond to a
change in the choice of the brain vacuum: the brain state
evolution or ``story'' is thus the story of the trade of the brain
with the surrounding world. The theory should then provide the
equations describing the brain evolution ``through the vacua'',
each vacuum for each instant of time of its history.

The brain evolution is thus similar to a time--ordered sequence of
photograms: each photogram represents the ``picture" of the brain
at a given instant of time. Putting together these photograms in
``temporal order" one gets a movie, i.e. the story (the evolution)
of open 



brain, which includes the brain--environment interaction
effects.

The evolution of a memory specified by a given code value, say
${\cal N}$, can be then represented as a trajectory of given
initial condition running over time--dependent vacumm states,
denoted by $|0(t)>_{{\cal N}}$, each one minimizing the free
energy functional (Pessa and Vitiello, 2003a, 2003b; Vitiello,
2003).
These trajectories are known (Manka, Kuczynski and Vitiello, 1986;
Del Giudice et al., 1988, Vitiello, 2003) to be {\it classical}
trajectories in the infinite volume limit: transition from one
representation to another inequivalent one would be strictly
forbidden in a quantum dynamics.

\bigskip
\bigskip
\section{Entanglement, chaos and coherence}

Since we have now two--modes (i.e. non--tilde and tilde modes),
the memory state $|0(t)>_{{\cal N}}$ turns out to be
a two-mode coherent state. This is known (Perelomov, 1986;
Vitiello 2003; Pessa and Vitiello, 2003a, 2003b) to be an
entangled state, i.e. it cannot be factorized into two
single--mode states, the non--tilde and the tilde one. The
physical meaning of such an entanglement between non-tilde and
tilde modes is in the fact that the brain dynamics is permanently
a dissipative dynamics. The entanglement, which is an unavoidable
mathematical result of dissipation, represents the impossibility
of cutting the links between the brain and the external world.

I remark that the entanglement is permanent in the large volume
limit. Due to boundary effects, however, a unitary transformation
could disentangle the tilde and non--tilde sectors: this may
result in a pathological state for the brain. It is known that
forced isolation of a subject produces pathological states of
various kinds.

I also observe that the tilde mode is not just a mathematical
fiction. It corresponds to a real excitation mode (quasiparticle)
of the brain arising as an effect of its interaction with the
environment: the couples of non--tilde/tilde dwq quanta represent
the correlation modes dynamically created in the brain as a
response to the brain--environment reciprocal influence. It is the
interaction between tilde and non--tilde modes that controls the
irreversible time evolution of the brain: these collective modes
are confined to live {\it in} the brain. They vanish as soon as
the links between the brain and the environment are cut.

Here, it is interesting to recall (Vitiello, 1998, 2001) that
structure and function constitute an un--dividable unity in the
frame of QFT: the dwq quanta are at the same time structure (they
are {\it real particles} confined to live inside the system) and
function, since they {\it are} the collective, macroscopic
correlations characterizing the brain functioning.

The structure/function unity in the dissipative model thus
accounts for the observed strong ``reciprocal dependence''
existing between the formation of neuronal correlates and nets and
the functional activity of the brain, including the brain's {\it
plasticity} and {\it adaptiveness}. The dissipative model implies
that the insurgence of some structural (physiological) pathologies
of the brain may be caused by the reduction and/or inhibition of
its functions due to externally imposed constraints in some severe
conditions.

As mentioned, transitions among unitary
inequivalent vacua may occur (phase transitions%
\index{phase transition}) for large but finite volume, due to
coupling with
the environment. Due to dissipation%
\index{dissipation} the brain appears as ``living over many
vacuum%
\index{vacuum state} states'' (continuously undergoing phase
transitions).
Even very weak (although above a certain threshold) perturbations
may drive the system through its macroscopic configurations.
In this way, occasional (random) weak perturbations play an
important r\^{o}le in the complex behavior of the brain activity.
In a recent paper (Pessa and Vitiello, 2003a, 2003b) the tilde
modes have been shown to be strictly related to the quantum noise
in the fluctuating random forces coupling the brain with the
environment.

It has been also found (Pessa and Vitiello, 2003a, 2003b) that,
under convenient conditions, in the infinite volume limit,
trajectories over the memory space are classical chaotic
trajectories (Hilborn, 1994), namely: i) ~they are bounded and
each trajectory does not intersect itself (trajectories are not
periodic); ii)~there are no intersections between trajectories
specified by different initial conditions; iii) ~trajectories of
different initial conditions are diverging trajectories.

In this connection, it is interesting to mention that some
experimental observations by Freeman (1990, 1996, 2000) show that
noisy fluctuations at the neuronal level may have a stabilizing
effect on brain activity, noise preventing to fall into some
unwanted state (attractor) and
being an essential ingredient for the neural chaotic%
\index{chaos} perceptual apparatus (especially in neural
aggregates of the olfactory system of laboratory animals).

In the dissipative model noise and chaos turn out to be natural
ingredients of the model. In particular, in the infinite volume
limit the chaotic behavior of the trajectories in memory space may
account for the high perceptive resolution in the recognition of
the perceptual inputs. Indeed, small differences in the codes
associated to external inputs may lead to diverging differences in
the corresponding memory paths. On the other side, it also happens
that codes differing only in a finite number of their components
(in the momentum space) may easily be recognized as being the
``same'' code, which makes possible that ``almost similar'' inputs
are recognized by the brain as ``equal'' inputs  (as in pattern
recognition).

Summarizing,  the brain may be viewed as a complex system with
(infinitely) many macroscopic configurations (the memory
states). Dissipation%
\index{dissipation} is recognized to be the root of such a
complexity.

The brain's many structural and dynamical levels
(the basic level of coherent condensation%
\index{condensation} of dwq%
\index{dipole wave quanta}, the cellular cytoskeleton%
\index{cytoskeleton} level, the neuronal dendritic level, and so
on) coexist, interact among themselves and influence each other's
functioning. The crucial point is that the different levels of
organization are not simply structural features of the brain;
their reciprocal interaction and their evolution is intrinsically
related to the basic quantum dissipative dynamics.

On the other hand, the brain's functional
stability is ensured by the system's ``coherent response%
\index{coherent response}'' to the multiplicity of external
stimuli. Thus dissipation also seems to suggest a solution to the
so called {\it binding problem}, namely the understanding of the
unitary response and behavior of apparently separated units and
physiological structures of the brain. In this connection see also
the holonomic theory by Pribram (1971, 1991).

The coherence properties of the memory states also explain how
memory remains stable and well protected within a highly excited
system, as indeed the brain is. Such a ``stability" is realized in
spite of the permanent electrochemical activity and the continual
response to external stimulation. The electrochemical activity
must also, of course, be coupled to the correlation modes which
are triggered by external stimuli. It is indeed the
electrochemical activity observed by neurophysiology that provides
a first response to external stimuli.

This has suggested (Stuart, Takahashi and Umezawa, 1978, 1979) to
model the memory mechanism as a separate mechanism from the
electrochemical processes of neuro-synaptic dynamics: the brain is
then a ``mixed" system involving two separate but interacting
levels. The memory level is a quantum dynamical level, the
electrochemical activity is at a classical level. The interaction
between the two dynamical levels is possible because the memory
state is a {\it macroscopic quantum state} due, indeed, to the
{\it coherence%
\index{coherence}} of the correlation modes. The coupling between
the quantum dynamical level and the classical electrochemical
level is then the coupling between two macroscopic entities. This
is analogous to the coupling
between classical acoustic waves and phonons%
\index{phonon} in crystals (phonons are the crystal NG quanta).
Such a coupling is possible since the macroscopic behavior of the
crystal ``resides'' in the phonon modes, so that the coupling
acoustic--waves/phonon is nothing but the coupling
acoustic--wave/crystal.

Finally, let me observe that, considering time--dependent
frequency for the dwq, modes with higher momentum are found to
possess longer life--time. Since the momentum is proportional to
the reciprocal of the distance over which the mode can propagate,
this means that modes with shorter range of propagation will
survive longer. On the contrary, modes with longer range of
propagation will decay sooner.  This mechanism may
produce the formation%
\index{domain formation} of ordered domains of finite different
sizes with different degree of stability: smaller domains would be
the more stable ones. Thus we arrive
at the dynamic formation%
\index{domain formation} of a hierarchy (according to their
life--time or equivalently to their sizes) of ordered domains
(Alfinito and Vitiello, 2000).
On the other hand, since any value of the momentum is in principle
allowed to the dwq, we also see that a scaling law is present in
the domain formation (any domain size is possible in view of the
momentum/size relation).

\bigskip
\bigskip
\section{The trade with the Double: a route to consciousness?}

We have seen that the tilde modes are a representation of the
environment ``modelled'' on the (non--tilde) system: they
constitute the time--reversed copy of it. And, we have seen, they
are ``necessary'', they cannot be eliminated from the game. The
mathematical operation of doubling the system degrees of freedom,
required by dissipation, thus turns out to produce the system's
{\it Double}. I have then suggested that consciousness mechanisms
might be involved in the continual ``trade'' (interaction) between
the non-tilde and the tilde modes (Vitiello, 1995, 2001).

Here I would be tempted to say: trade ``between the subject and
his Double''. However, the word ``subject'' may be evocative of
rich but intricate philosophical scenarios, which here are
absolutely out of my considerations. I have experienced indeed
that using in a simple minded way that word in connection with the
brain (as I did in my book (Vitiello, 2001)) may be highly
misleading, pushing the reader far from the much more modest, but
concrete, mathematical and physical features of the dissipative
model.

My attention is rather on the dynamics, the ``inter--action'', the
trade, the ``between'' (as Gordon Globus (2003) would say),
l'``entre--deux'' (as Nadia Prete (2003) would prefer). The use of
the word ``subject'' could instead evoke the idea of something,
the ``one'' (the non--tilde), pre-existing the relation with the
other ``one'', his Double (tilde). However, this would correspond
neither to the physics, nor to the mathematics, both of which are
my fixed starting points.

The physics of the problem is, technically speaking, a
non--perturbative physics, the one of the open systems, and it can
be shown that in such a case the system--environment
``inter--action'' cannot be switched off (Celeghini, Rasetti and
Vitiello, 1992). This means that we cannot even think of the
system deprived of its physical essence which is its openness
(even the physiology tells us that an isolated brain is a dead
brain; namely, if ``closed'', it does not exist {\it as a brain}).
The physics does not allow the existence of the ``one''
(non--tilde) {\it independently} of the existence of the ``other
one'' (tilde), and vice versa.

The mathematics, on the other hand, imposes a strong limit on the
description (the ``language'') we have to use for a quantum
dissipative system: we cannot avoid starting from two reciprocal
(in the mirror of time) images. This ``un--divided two'',
mathematics tells us, is more elementary than ``the one''. The
non--tilde one ``cannot'' be the subject.

The temptation could be to think that the Double is the subject.
But this simply means being captured by the Narcissus
self-mirroring fatal trap (Vitiello, 2001): it is equivalent to
think of the brain as the subject, and vice versa in an endless
loop. The Double is not trivially the system image. It is the
environment representation {\it modelled} on the system. The
Double cannot be the subject.

Tilde and non--tilde cannot {\it individually} pre--exist prior to
their being each other's images, an ``un-dividable two''. They are
actors {\it forced} (without alternative choice) to be on the
stage. The ``one'', the subject, is the action, the play, their
entre--deux. This is the meaning of the entanglement: the
entangled state cannot be factorized (is un-dividable) into two
single-mode states. Non--tilde and tilde modes share a common,
entangled vacuum at each instant of time.

In some sense, here we face the root of the (ontological)
prejudice that some ``being'' might exist as a ``closed'', i.e.
non--interacting, system, and therefore, capable to exist {\it by
itself}, {\it independently} of the existence of any other system,
{\it complete} in its own individuality (Vitiello, 1997, 2001). If
so, it might also happen that such a system could be, in absolute,
the {\it only} existent system (``being'').

Such a prejudice seems to be intrinsic to our same language, where
any action presupposes pre--existing actors having the possibility
of being fully non--interacting, and thus each one {\it
independently} existing from the other one (fully disentangled)
{\it before} the action started. Notable exceptions (Stamenov,
2001) might be those actions, such as {\it to exchange}, {\it to
trade}, indeed, which exclude the {\it separate} (disentangled)
pre--existence of the actors: to be possible that such actions
could occur, the {\it joint} existence of (at least) a couple of
actors is necessary (even if not sufficient, of course). Thus,
those actions are special ones in that they presuppose entangled
existences of the actors. Each one of these cannot exist by
himself. And also, actors cannot be {\it separated} from their
action and vice versa. Without exchange there are no exchangers,
and vice versa. Such a situation also reminds me of the rheomode
language of Bohm, whose structure is aimed to allow ``the verb
rather than the noun to play a primary role'' (Bohm, 1980; cf.
Stamenov, 2003 for a discussion on Bohm's rheomode language).

The unavoidable dialog with the Double is the continual,
changeable and reciprocal (non-linear) interaction with the
environment. If the consciousness phenomenon basically resides in
such a permanent dialog, one of its characterizations seems to be
the relational (medial) one, which agrees with Desideri's
standpoint (Desideri, 2003). Consciousness seems thus to be rooted
and diffused in the large brain--environment world, in the
dissipative brain dynamics. There is no conflict between the {\it
subjectiveness} of the first person experience of consciousness
and the {\it objectiveness} of the external world
{\footnote{Although it might sound philosophically unpleasant, I
adopt the physicist's working hypothesis that the external world
is objectively existing. In rough words, this amounts to adopt the
working hypothesis 
that we do exchange energy with some other system. For example, 
we do need to eat. Without eating
we cannot think. Of course, this does not mean that thinking is
less important than eating, but simply that neglecting to eat
leads to weak (or null) thinking.}}. Without such an {\it objectiveness}
there would be no possibility of ``openness'' (openness on what?),
no dissipation out of which consciousness could arise.
Objectiveness of the external world is the primary, necessary
condition for consciousness to exist.

On the other hand, the question Desideri poses, namely ``whether
it is possible to reverse also the relationship between structure
and function and then if it is possible to consider brain as a
function of consciousness''(Desideri, 2003) also finds a positive
answer in the dissipative quantum model. The answer is positive in
a true physical sense, since the brain cannot avoid to be an
active/passive system, and promoting or inhibiting its activity
(summing up in the consciousness) would produce the creation or
destruction, respectively, of structural features of the brain,
such as, e.g., long range correlations, pattern structures. As
observed in Section 4, the different levels of organization are
not simply structural features of the brain, their reciprocal
interaction and their evolution is intrinsically related to the
brain--environment entanglement, namely to that medial ``one''
which is the dialog with the Double. In this sense, the
adaptiveness, the plasticity of the brain {\it is} the function of
consciousness.



It is also interesting to observe that the dialog with the Double
is ``evolutive'' and never repeats itself in the same form: from
one side, it carries the memory, the story of the past; from the
other side, the permanent openness on the world implies its
continual updating. Recurrent resolutions into ''new synthesis''
of the non--tilde/tilde reciprocal presence are thus reached. The
mentioned process of minimizing the free energy, namely of
reaching the equilibrium between the numbers of non--tilde and
tilde modes, is indeed the process by which such synthesis are
recurrently reached by permanently tuning the constantly renewed
brain--environment ''relation''. The actors are never engaged in a
boring reply. And such a {\it truly dialectic} relation with the
Double is inserted in the unidirectional flow of time, it is
itself a ``witness'' of the flow of time. This depends on the fact
that its mathematical description is provided by the coupling term
in the time evolution operator and this is proportional to the
entropy operator. It is possible to talk of {\it unidirectional}
flow of time because time--reversal symmetry is broken due to
dissipation (cf. also the beginning of Section 2). Then, the time
axis gets divided by a {\it singular} point: {\it the origin},
which divides the past from the future. The singularity of this
point consists in the fact that {\it it cannot be translated}, it
is the {\it Now}.

Without dissipation, {\it any} point, any time, can be {\it
arbitrarily} taken to be the origin of the time axis, which means
that the origin (and any other point on the time axis) can be
freely {\it translated} without inducing any observable change in
the system (time translational invariance): thus there is no {\it
singular} origin of the time. There is no {\it Now}. All the
origins are alike. There is no a {\it true} origin.

In the absence of dissipation, we could say that time, in its
flowing, swallows those fictitious Nows we might assign as
(non-singular) origins on its axis, as $K\rho \acute{o} \nu o
\varsigma$ eats his sons. This destructive property of time
(oblivion) is, paradoxically, eluded, avoided by dissipation:
dissipation introduces a life--time, a time scale which carries
the {\it memory} of ``when'' (the origin) the dissipative system
``has started''. From the observation of a (radiative) decay
process (typically with carbon fourteen) we ``can trace back'' the
time, reach the origin and say ``how old'' is the object we are
interested in. So we know where the true, non--forgettable origin
(the truth), not a fictitious, false one easily eaten by $K\rho
\acute{o} \nu o \varsigma$, sits on the time axis. Memory
(non--oblivion) and truth are the same thing, which the ancient
Greeks denoted, indeed, with the same word,
$\alpha\lambda\acute{\eta}\theta\epsilon\iota\alpha$ (Tagliagambe,
1995; Vitiello, 1997, 2001).

The Now is that point on the time--mirror where the non--tilde and
the tilde, reciprocal time--reversed images, join together, in the
{\it present} (Vitiello, 1997, 2001). The non--tilde {\it unveils}
its Double and they conjugate  in a circular (non--linear)
recognition, each being ``exposed'' to the other's eyes. Perhaps
this is {\it intuition}, the {\it instantaneous apprehension}
(Webster Dictionary, 1968) of the ``between''. Literally, {\it
intueri} is such a looking inside ``without the conscious use of
reasoning'' (Webster Dictionary, 1968), an immediate, out of time,
not in the past, not in the future, act of {\it unconscious}
knowledge, an ``unknowable act'' (Plotnitsky, 2002) of knowledge.
An act which repeats itself continuously, not translating the Now
(dissipation forbids it!), but re--creating another independent,
but equally true, Now, in an endless, dense sequence of Nows, all
different, singular origins of different paths in the future, all
starting points of chaotic memory paths in the memory space, which
then we recognize as the ``identity'' or the ``self'' space.
Identity, dynamically living in the memory space through the
dissipative Nows, thus escapes the destructive fury of $K\rho
\acute{o} \nu o \varsigma$.

Perhaps, in these Nows is realized the primary property of
consciousness, the one of self--questioning (Desideri, 2003), i.e.
the unveiling the Double, and the photographer's
``sur-prise''...``when at the precise instant an image suddenly
stands out and the eye stops'' forcing ``the time to stop his
course''  (Prete, 2003): ``and suddenly, all at once, the veil is
torn away, I have understood, I have {\it seen}''(Sartre, 1990;
see also the related discussion by Prete, 2003).

Unveiling the Double is then to see and to be seen, the
$\sigma\upsilon\nu\epsilon\iota\delta\acute{\omega}\varsigma$, the
being conscious of the ancient Greeks, which literally is to ``see
together'', indeed; or, as in the lifting the veil in the Prete's
photobjects, `` more precisely, to have a perception of this
togetherness as a whole and to understand that it was made of two
images in strong relation'' (Prete, 2003); or else Bohm's
self-recursive mirroring loops of the spontaneous and unrestricted
act of ``lifting into attention'' (Bohm, 1980; Stamenov, 2003):
$\sigma\upsilon\nu\epsilon\iota\delta\acute{\omega}\varsigma$ then
comes to be confidants, secret friends (Bandini, 2002), to be each
other ``witness''.

Such a sudden act of knowledge remains, however, an intuitive
knowledge, an {\it unum}, not susceptible to be ``divided'' into
rational steps, thinkable but ``non--computational'', not
``translatable'' into a language (i.e. logical) frame, which would
require its breaking up (analysis) into linguistic fragments (cf.
the traditional language fragmentation discussed in Bhom, 1980,
and the related discussion by Stamenov, 2003). (It is interesting
that the $\epsilon\iota\delta\acute{\omega}\varsigma$ in the word
$\sigma\upsilon\nu\epsilon\iota\delta\acute{\omega}\varsigma$
(being conscious) denotes the act of immediate vision; the word
$o\rho\acute{\alpha}\omega$ is used instead for the act of lasting
vision (Bonazzi, 1936)).

In conclusion, from the sequence of these acts inserted into the
``objective'' time flow a sequence of independent, subjective Nows
is generated, which constitute the multi--time dimensions of the
{\it self}, its {\it own} time space, the {\it dynamic} archive of
chaotic trajectories in the memory space which depicts its
identity; that spring of time--lines through which the self can
move ``freely'', apparently unconstrained by the external
time--ordering.

Without such an internal freedom there could be neither the
``pleasure'' of the perception (the
$\alpha\acute{\iota}\sigma\theta\eta\sigma\iota\varsigma$), the
aesthetical dimension, that erotic charge of the {\it unveiling},
which continuously renews itself in the dialogic relation with the
Double, nor the ``active response'' to the world. Neither
pleasure, nor intentionality could be allowed in a rigidly
constrained system. Active responses imply responsibility and thus
they become moral, ethical responses through which the self and
its Double become part of the larger social dialog. Aesthetical
pleasure unavoidably implies disclosure, to {\it manifest}
``signs'', artistic {\it communication}. An interpersonal,
collective level of consciousness then arises, a larger stage
where again the actors are mutually dependent, each one bounded
(entangled) in his very existence (including any sort of physical
needs) to the other ones, simply non-existing without the others.

\bigskip
\bigskip
\section{Doubts and mistakes. Toward the construction of an erratic device.}

I finally observe that the strong influence of even slight changes
in the initial conditions on the memory paths (their chaotic
behavior) leads us to consider the r\^{o}le of the  ``doubt'' in
consciousness mechanisms (Desideri, 1998). In this connection I
will also very briefly comment on a provocative proposal of mine:
to construct an artificial device able to make ``mistakes'',
namely able of taking a step, in its behavior, not logically
consequent from the previous ones, or not belonging to any
pre--ordered chain of steps or events, an erratic step. For
shortness, and in a provisional way, I will refer to it as to the
``erratic device''. Such a device is perhaps in strict relation
with an artificial conscious device (if ever it will be possible
to construct an artificial conscious device!).

My erratic device is not a machine ``out of order'', not properly
functioning. It {\it cannot} be a machine at all, since a machine,
in the usual sense, is {\it by definition} (and by construction)
something which must work {\it properly}, in a strictly predictive
way, producing processes of sequentially ordered steps according
to some functional logic. Also, the erratic device is not meant to
be a device exhibiting chaotic behavior: the {\it value} (!) of
the mistake is in its infrequent occurrence, an exceptional
``novelty'' with respect to an otherwise ``normal'' (correct)
behavior.

But let me go back to the dissipative quantum model of brain.
There, tilde modes also account for the quantum noise in the
fluctuating forces coupling the brain with the environment. The
dialog with the Double lives therefore on a noisy background of
quantum fluctuations. ``Listening'' sometimes at such noisy
background in the continual dialog (self--questioning) might
slightly perturb the initial conditions of the memory paths and
manifest in their drastic differences. This might be sometimes a
welcome event, pushing the brain activity out of unwanted loops or
fixations (attractors), (which also suggests a possible relation
with Freeman's (1990, 1996, 2000) observations on neuronal noisy
activity). Doubt might well be such a kind of self--questioning in
a noisy background, being tempted by new perspectives, testing new
standpoints by more or less slightly perturbing old certainties,
leaving room for erratic fluctuations, listening to them; in a
word, allowing fuzziness in the initial conditions, the starting
assumptions of our travelling in the memory space (our archive of
certainties); the consequences of the doubt will be then
chaotically diverging trajectories in such a space. Consciousness
modes then acquire their uncertain (doubtful) predictability with
their precious unfaithfulness, their secret flavor of
subjectivity, their full {\it autonomy}.

I suspect that the great privilege of being able of making
mistakes finds its roots in these consciousness features. And
perhaps here is the bridge between the program of constructing the
erratic device and the one of constructing an artificial conscious
device.

Perhaps, if ever it will be possible to construct a conscious
artificial device, it will not be indeed a ``machine'', i.e. its
behavior cannot be like a chain of logically predetermined steps,
it must be an artificial being taking upon itself the best of the
human model: unpredictably erratic, able to learn, but unfaithful,
full of doubts,  fully entangled to the world, but irreducibly
free. We might name it Spartacus.

\smallskip
\bigskip

\newpage



\Large{\bf References}

\normalsize

\begin{list}{}{}

\item
Alfinito, ~E. and G.~Vitiello (2000). Formation and life--time of
memory domains in the dissipative quantum model of brain, {\it
Int. J. Mod. Phys.} {\bf B14}, 853-868.

\item  
Bandini, ~F. (2002). La Coscienza e il Tempo. in {\it Exploring
consciousness}, p. 147--153. Milano: Fondazione Carlo Erba.

\item
Bonazzi, ~B. (1936). {\it Dizionario Greco--Italiano}. Napoli: A.
Morano.

\item
Bohm, ~D. (1980). {\it Wholeness and the implicate order}.
London:Routledge and Kegan Paul.

\item 
Celeghini, ~E., M.~Rasetti and G.~Vitiello (1992). Quantum
Dissipation, {\it Ann. Phys.} {\bf 215}, 156-170.

\item  
Del Giudice, ~E., S. ~Doglia, M. ~Milani and G. ~Vitiello (1985).
A quantum field theoretical approach to the collective behavior of
biological systems. {\it Nucl. Phys. B251 [FS 13]}, 375-400.

\item  
Del Giudice, ~E., S. ~Doglia, M. ~Milani and G. ~Vitiello (1986).
Electromagnetic field and spontaneous symmetry breakdown in
biological matter. {\it Nucl. Phys. B275 [FS 17]}, 185-199.

\item  
Del Giudice, ~E., R.~Manka, M.~Milani and G.~Vitiello (1988).
Non-constant order parameter and vacuum evolution, {\it
Phys.Lett.\ } {\bf A 206}, 661-664.

\item  
Desideri, ~F. (1998). {\it L'ascolto della coscienza}. Milano:
Feltrinelli.

\item  
Desideri, ~F.  (2003). The self-transcendence of consciousness
towards its models. {\it This Volume}. Amsterdam: John Benjamins.

\item  
Freeman, ~W.~J. (1990). On the the problem of
anomalous dispersion in chaotic%
\index{chaos} phase transitions%
\index{phase transition} of neural masses, and its significance
for the management of perceptual information in brains, in H.
~Haken and M. ~Stadler (Eds.), {\it Synergetics of cognition} {\bf
45}, 126--143. Springer Verlag, Berlin.

\item  
Freeman, ~W.~J. (1996). Random activity at the microscopic neural
level in cortex (''noise'') sustains and is regulated by
low dimensional dynamics of macroscopic cortical activity (''chaos%
\index{chaos}''), {\it Intern. J. of Neural Systems} {\bf 7},
473--480.

\item  
Freeman, ~W.~J. (2000). {\it Neurodynamics: An exploration of
mesoscopic brain dynamics}. Springer, Berlin.

\item  
Globus, ~G. (2003). {\it Quantum closures and disclosures:
Thinking--togheter postphenomenology and quantum brain dynamics}.
Amsterdam: John Benjamins.

\item  
Hilborn, ~R. (1994). {\it Chaos and nonlinear Dynamics}. Oxford:
University Press.

\item  
Jibu, ~M., and ~K. ~Yasue (1995). {\it Quantum brain
dynamics and consciousness%
\index{consciousness}}. Amsterdam: John Benjamins.


\item  
Manka, ~R., J.~Kuczynski and G.~Vitiello (1986). Vacuum structure
and temperature effects. {\it Nucl. Phys.\ } {\bf B 276},
533--548.

\item
Ojima, ~I. (1981). Gauge fields at finite temperature -- ''Thermo
Field Dynamics'' and the KMS condition and their extension to
gauge theories. {\it Annals of Physics\ } {\bf 137}, 1--37.

\item  
Perelomov, ~A. (1986). {\it Generalized Coherent States and Their
Applications}. Berlin: Springer.

\item  
Pessa, ~E. and G.~Vitiello (2003a). Quantum noise, entanglement
and chaos in the quantum field theory of mind/brain states. {\it
Mind and Matter} {\bf 1}, 59--79. arXiv:q-bio.OT/0309009.

\item  
Pessa, ~E. and G.~Vitiello (2003b). Quantum noise induced
entanglement and chaos in the dissipative quantum model of brain.

\item  
Plotnitsky, ~A. (2002). {\it The knowable and the unknowable}. Ann
Arbor: The University of Michigan.

\item  
Prete, ~N. (2003). Doubling image to face the obscenity of
photography. {\it This Volume}. Amsterdam: John Benjamins.

\item  
Pribram, ~K.~H. (1971). {\it Languages of the brain}. Englewood
Cliffs, N.J.: Prentice-Hall.

\item  
Pribram, ~K.~H.  (1991). {\it Brain and perception}.  Hillsdale,
N. J.: Lawrence Erlbaum.

\item  
Ricciardi, ~L.~M. and H.~Umezawa (1967). Brain and physics of
many-body problems, {\it Kybernetik} {\bf 4}, 44--48. {\it Reprint
in this Volume}.

\item  
Sartre, ~J.~P. (1990). {\it La nausea}. Torino: Giulio Einaudi.
page 171.

\item  
Stamenov, ~M. ~I. (2001). Nouns, Verbs, and Consciousness.
{Pre--Conference Workshop}, {\it Toward a science of
consciousness. Consciousness and its place in Nature}. Sk\"{o}vde,
Sweden: unpublished.

\item  
Stamenov, ~M. ~I. (2003). The rheomode of language of David Bohm
as a way to re-construct the access to physical reality. {\it This
Volume}.

\item  
Stuart, ~C.~I.~J., Y.~Takahashi and H.~Umezawa (1978). On the
stability and non-local properties of memory, {\it J. Theor.
Biol.} {\bf 71}, 605--618.

\item  
Stuart, ~C.~I.~J.,  Y.~Takahashi and H.~Umezawa (1979). Mixed
system brain dynamics: neural memory as a macroscopic ordered
state, {\it Found. Phys.} {\bf 9}, 301--327.

\item  
Tagliagambe, ~S. (1995). Creativita'. {\it Atque 12}%
, 25-46.

\item  
Takahashi, ~Y. and H. ~Umezawa (1975). Thermo%
\index{thermo field dynamics} field dynamics. {\it Collective
Phenomena 2}, 55-80.

\item
Titus Lucretius Carus (99--55 B.C.). {\it De rerum natura}.
Translation by Walter Englert, 2003. Newbury Port, Ma: Focus.

\item  
Umezawa, ~H. (1993). {\it Advanced field theory: micro, macro and
thermal concepts}. New York: American Institute of Physics.

\item  
Vitiello, ~G. (1995). Dissipation and memory capacity in the
quantum brain model, {\it Int. J. Mod. Phys. } {\bf B9}, 973--989.

\item  
Vitiello, ~G. (1997).
Dissipazione e coscienza. {\it %
Atque 16}, 171--198.

\item  
Vitiello, ~G. (1998). Structure and function. An open letter to
Patricia Churchland. In S.R. Hameroff, A. W. Kaszniak and A. C.
Scott (Eds.), {\it Toward a science of consciousness II. The
second Tucson Discussions and debates} p. 231--236. Cambridge: MIT
Press.

\item  
Vitiello, ~G. (2001). {\it My Double Unveiled}. Amsterdam: John
Benjamins.

\item  
Vitiello, ~G. (2003). Classical chaotic trajectories in quantum
field theory, arXiv:hep-th/0309197.

\item  
Webster's New World Dictionary (1968). College Edition, Cleveland:
The World Pu. Co..


\end{list}

\end{document}